# Gas Electron Multiplier detectors with high reliability and stability

B.M.Ovchinnikov[1], V.V.Parusov[1] and Yu.B.Ovchinnikov[2]

[1]Institute for Nuclear Research of Russian Academy of Sciences, Moscow, Russia

[2]National Physical Laboratory, Teddington, Middlesex, TW11 0LW, UK

# Abstract

The Gas Electron Multiplier detectors with wire and metallic electrodes, with a gas filling in the gap between them were proposed and tested. The main advantage of these Gas Electron Multipliers compared to standard ones consists in their increased stability and reliability. The experimental results on testing of such detectors with gaps between the electrodes of 1 and 3 mm are reported. It is demonstrated, that the best gas filling for the gas electron multipliers is neon with small admixture of quenching gases (for example, ($N_2$+$H_2O$) at≈100ppm). This filling offers the greatest coefficient of proportional multiplication as compared with other gases, at small electric potential difference between the GEM electrodes, in absence of streamer discharges in the proportional region. The results on operation of the multi-channel gas electron multiplier with wire cathode and continuous anode filled with *Ne, Ar, Ar+$CH_4$* and *Ar+1%Xe* are presented also. Based on the experimental observations, the explanation of the mechanism of formation of streamers is given.

E-mail address: ovchin@inr.ru

# Introduction

GEM detectors have good position resolution on *x*-, *y*- coordinates and may provide high resolution on *z*-coordinate also owing to extraction of electron components of avalanches [1]. The other advantage of GEM detectors is a high suppression of secondary avalanches production on electrodes of a chamber by photons from avalanches in GEM holes. At the same time the secondary avalanche is produced in gas phase after every primary avalanche for Penning mixtures [1]. The primary avalanches are produced by ionization electrons from detected event and the secondary avalanches are produced by photons from primary avalanches on component of mixture with low ionization potential. The tails of these secondary avalanches have continuous character and exceed considerably the primary avalanches in duration, so the signals from them can be removed easily by differentiation.

The Gas Electron Multipliers, originally proposed by Sauli in 1997 [2], are developed very intensively over last years. These GEM are produced by perforation of plastic plates, both sides of which are covered with metallic or resistive coatings. Considerable shortcomings of these plastic GEM are their low reliability and stability. The streamer and spark discharges and also the streams of positive ions from the proportional avalanches disperse the metallic and resistive coatings of the GEM electrodes with subsequent deposition of them on the walls of the GEM holes, anode and the chamber insulators with the following current leakages and discharges. The accidental spark and streamer discharges over the plastic surface create also the conductive channels with following discharges on them.

The GEM with wire and metallic electrodes, which were proposed, produced and tested in works [3-6], are free of such shortcomings. It supposes, that the chamber insulators are mechanically protected from the deposition of molecules on them.

The stability of plastic GEM is limited by accumulation of the electrical charges on the plastic walls of the holes, which is not the case of GEM with wire and metallic electrodes.

.

# A multi-channel wire gas electron multiplier with a gap of 1mm

The design of the multi-channel wire gas electron multiplier (MWGEM) is shown in Fig.1. The electrodes of MWGEM have rectangular openings of 0.5mm x 0.5mm with 1.5mm steps between them in two orthogonal directions. The gap between the electrodes is equal to 1mm, while the total working area has a diameter of 20mm. The gas multiplication of electrons in the MWGEM happens between the rectangular openings of the electrodes.

The tests of MWGEM were conducted in a chamber containing a cathode C, MWGEM and anode A (Fig.2). The gap between the cathode C and the MWGEM is 13mm and a gap between the MWGEM and the anode A is 6mm.

The chamber was filled with pure commercial neon at pressure of 1bar. The space in the gap between the cathode and the MWGEM was irradiated by $\alpha(Pu^{239})$ or $\beta(Ni^{63})$ particles.

The signals from the anode are amplified by a charge sensitive amplifier BUS 2-96 and recorded with a memorizing oscilloscope.

At the irradiation of the chamber by $\beta$-particles *($Ni^{63}$, ~100 $\beta$/s)* the coefficient of proportional multiplication up to $8\cdot10^3$ was obtained (Fig.3a). The streamer discharges are practically absent for the gain $K_{ampl}<8\cdot10^3$, while above that gain value the signals consist mostly from the streamer discharges.

The maximal coefficient of proportional multiplication of about 300 was obtained at the irradiation of the chamber by $\alpha$-particles *($Pu^{239}$)* (Fig.3b).

The commercial neon, which was used in this experiment, is contaminated with other gases, like $H_2O$, $N_2$, $O_2$, at concentrations of ≤2ppm. In addition, as far as the chamber of the MWGEM was not baked for cleaning of its walls from the air and water admixtures, the neon in the chamber contained also quenching admixtures of ($N_2+O_2+H_2O$) at the level of 100ppm. As it was shown earlier in [1], the presence of these admixtures in neon leads to formation of secondary Penning avalanches, which have duration of 20–100μs. On the other hand, the amplitudes of these secondary avalanches, for the concentration of contaminations in neon at the level of about 100ppm, are small in the whole range of proportional signals and they do not produce streamer discharges. Only near the maximal coefficient of amplification, $K_{ampl}^{\max}$, the secondary avalanches increase and turn into streamers.

The smaller maximal amplification for $\alpha$-particles, compared to $\beta$-particles, $K_{ampl}^{\max}(\alpha)<K_{ampl}^{\max}(\beta)$, can be explained by larger amplitudes of secondary avalanches for $\alpha$-particles.
It is already known, that increase of secondary avalanches in mixtures of noble gases with quenching additions (5-20)%, compared to noble gases, containing ($H_2O+N_2+O_2$) at ≈100ppm leads to decrease of the maximum amplification $K_{ampl}^{\max}$. In the work [7] it was demonstrated that the maximal amplification in the $Ne+5\%CH_4$ mixture is smaller than in $Ne+((H_2O+N_2)$ at≈100ppm). In [8] it is convincingly shown, that increasing of the quenching additions for the mixtures *Ne+DME, Ar+DME, Ar*+izobutylene decreases the maximal amplification. In the work [6] the amplification of $K_{ampl}^{\max}=100$ for the $Ar+20\%CH_4$ mixture has been achieved, while for

($Ar+(H_2O+N_2)$≈100ppm) it was equal to $K_{ampl}^{\max}=10^3$.

The main purpose of the quenching additions in MWPC is to remove the photon feedback on its electrodes. On the other hand, the photon feedback in GEM is small, because of low photon flux to cathode, which is restricted by angles. Therefore, small amounts (~100ppm) of quenching gases admixtures (for example $H_2O+N_2$) are enough to suppress the photon feedback in GEM.

At the irradiation of the MWGEM filled with ($Ne+(H_2O+N_2)$≈100ppm), at pressure of 0.4bar by $\beta$-particles the maximal coefficient of proportional multiplication of $10^5$ was obtained with an additional multiplication in the anode gap of $K_{ampl}^{anod}=14$ (Fig.10c).

To compensate the electrostatic attraction of the electrodes of MWGEM of large area, the dielectric spacers between the electrodes can be used. To decrease leakage currents and the discharges, which may take place due to deposition on the spacers of the electrode's material, these spacers must be made of dielectric material with enhanced surface area.

# Multi-channel wire gas electron multiplier with wire cathode and continuous anode (MWCAT)

The prototype of GEM [2] and MICROMEGAS [9] is a CAT [10], which consists of a net-cathode and a continuous anode, with a gas gap between them of 0.1-1.0mm. The MWCAT detector with the same configuration of its winded electrode as in the MWGEM, have been investigated in this work. Tests of the MWCAT with a gap of 1mm have been conducted in a chamber (Fig.4) with the gap between the cathode C of the chamber and the MWCAT equal to 13mm.

The dependence of the gain $K_{ampl}$ = f($V_d$) was obtained (Fig.3c) at the filling of the chamber with a $Ne+((H_2O+N_2)$ at≈100ppm) and at irradiation by $\beta$-particles. In that case, the maximal coefficient of proportional amplification is about $1.5\cdot10^4$. There are only rare streamers at the gain of $K_{ampl}=10^4$. For the higher gain the rapid increase of the number of streamer discharges is detected. Therefore, a sharp transition from an area of proportional multiplication to a streamer discharges take place for the MWCAT detector.

The results of measurements for a chamber filled with Penning mixture $Ar+20\%\ CH_4$ (1bar) and irradiated by $\beta$-particles are presented in fig.5. The maximal coefficient of proportional multiplication of about 100 is observed (curve a). In that case, the streamer discharges (curve b) are observed for the whole range of proportional signals. The amplitude of streamer discharges stops to increase at the voltage on a resistive divider of $V_d$ ~3kV, that can be explained by the slowing of the increase of the potential difference between the MWCAT electrodes (Fig.5c) due to increasing of the number of streamer discharges.

A typical shape of the streamer signals for $Ar+20\%\ CH_4$ gas mixture is presented in fig.6. The large quantity of the streamers, as well as their long time duration, in this gas mixture can be explained by the formation after every primary avalanche of large secondary avalanche [1]. Near $K_{ampl}^{max}$ all secondary avalanches turn to streamers.

Fig.7 shows the results of measurements performed for the chamber filled with commercial argon. As before, the chamber was not baked prior to these measurements. The maximal coefficient of proportional multiplication of electrons up to $10^3$ (curve a) has been observed, which is by an order of the magnitude greater, compared to the $Ar+20\%\ CH_4$ gas mixture. The streamer discharges are observed in the whole range of proportional signals, the amplitude of which is saturated at $V_d$ ~2kV (curve b).

The results of measurements at filling of the chamber with mixture $Ar+1\%\ Xe$ (P=1bar) are presented in Fig.8. In that case, the streamer discharges with frequency of about 1 Hz are observed within the whole range of proportional amplification.

# The MWGEM with a gap of 3mm

The operation of MWGEM with a gap of 3mm, filled with a $Ne+((H_2O+N_2)$ at≈ 100ppm) at pressures of 0.4 and 1.0bar was investigated. The structure of MWGEM electrodes and the chamber for MWGEM testing are the same as for MWGEM with a gap of 1mm (Fig.1, 2).

The dependences obtained for coefficients of proportional electron multiplication from potential difference between MWGEM electrodes for a gap of 3mm are shown in Fig.9. The field strength in the anode gap for these measurements was less than the threshold of electron multiplication, $E_A^{thr}$ =420V/cm (for P=0.4bar) and $E_A^{thr}$ =500V/cm (for P=1bar).

The increasing of the MWGEM gap from 1 to 3mm leads to increase of the maximal amplification, $K_{ampl}^{max}$, by factor of 10 by expense of increasing of the avalanche multiplication path (see Fig.10c (1mm) and Fig.10b (3mm)). For the gap of 3mm, the range of proportional multiplication without streamers is also increased up to one order of the magnitude.

# MWGEM with additional multiplication in anode gap

To increase the amplification and the working range of the MWGEM, the field strength in the anode gap was increased to 520V·cm$^{-1}$. This provided additional amplification $K_{ampl}^{anod}$=14 to the main amplification $K_{ampl}$(MWGEM).

The coefficient $K_{ampl}^{anod}$ =14 was measured by irradiation the drift gap of the MWGEM with $\alpha$-particles and 80V potential difference between the MWGEM electrodes. This voltage was not strong enough for electron multiplication, but sufficient to transfer electrons through the MWGEM to the anode gap.

The results for the simultaneous MWGEM and anode gap proportional amplification ($K_{ampl}^{tot}$=$K_{ampl}$(MWGEM) × $K_{ampl}^{anod}$), for the gap widths of 1 and 3mm and neon pressure of 0.4bar, are shown in Fig.10.

Comparing the results for the MWGEM without electrons multiplication in the anode gap (Fig.9):

[$\beta^-$, 3mm, 0.4bar, $K_{ampl}^{max}$(MWGEM) =$10^5$ (50% streamers)] and

[$\alpha$, 3mm, 0.4bar, $K_{ampl}^{max}$(MWGEM) =$6\cdot10^3$ (20% streamers)],

to the MWGEM with the additional electrons multiplication in anode gap (Fig.10):

[$\beta^-$, 3mm, 0.4bar, $K_{tot}$= $K_{ampl}^{max}$(MWGEM) × $K_{ampl}^{anod}$ =$1.4\cdot10^5\times14=2\cdot10^6$ (20% streamers)] and

[$\alpha$, 3mm, 0.4bar, $K_{tot}$= $K_{ampl}^{max}$(MWGEM) × $K_{ampl}^{anod}$ =$8\cdot10^3\times14=1.2\cdot10^5$ (50% streamers)],

it may be concluded, that the additional multiplication in the anode gap does not influence the multiplication of the MWGEM, but increases the total range of proportional multiplication (without streamers) by 14 times.

# Multi-channel gas electron multiplier with metallic electrodes

The MWGEM, filled with $(Ne+(H_2O+N_2)$≈100ppm), have high reliability, stability and rather high coefficients of amplification $K_{ampl}^{max}$=$10^5$-$10^6$, without streamers in the range of proportional amplification.

One of the problems of the MWGEM is a noise caused by the microphone effect, which takes place in presence of mechanical vibrations. In addition, the MWGEM makes difficult to design devices of complex shape (for example, of cylindrical or spherical shape).

The design of the multi-channel gas electron multiplier (MGEM) with metallic electrodes is shown in Fig.11. The electrodes of MGEM are produced from brass plates with thickness of 1mm, round holes of 1mm in diameter and placed at a distance of 1.5mm from each other. The gap between the two electrodes is equal to 3mm, while the total working area has a diameter of 20mm.

The tests of MGEM were conducted in the chamber Fig.2, filled with *(Ne+(*$H_2O+N_2$*)≈*100ppm).

The results of the MGEM testing at neon pressure of 1bar are shown in Fig.12.The chamber was not backed in these measurements.

The electrons from MGEM are effectively extracted in the anode gap only when the potential difference in the gap ≥420V. At the anode voltage of 420V the electrons in anode gap are multiplied with the coefficient $K_{ampl}^{anod}$ =10. The reason for such a bad permeability of the lower electrode of the MGEM consists in a low ratio between the diameter of the holes and the thickness of the electrodes (1:1). For improvement of the electrode permeability it is necessary to increase the hole diameter of electrodes up to 1.5-2.0mm and to decrease the electrodes thickness down to 0.5mm.

In Fig.13 the dependences of the total proportional electron multiplication coefficients $K_{ampl}^{tot} = K_{ampl}(\text{MGEM}) \times K_{ampl}^{anod}$ on the potential difference between the MGEM electrodes, at gas filling pressure of 0.4bar and in presence of amplification of electrons in the anode gap, are shown:

1. $V_a$=250V ( $K_{ampl}^{anod}$ =1),
2. $V_a$=300V ( $K_{ampl}^{anod}$ =6),
3. $V_a$=320V ( $K_{ampl}^{anod}$ =15).

# Investigations of the influence of $H_2O$ and $N_2$ micro admixtures to the operation of MGEM

The shapes of signals of the MGEM, taken for different content of gas fillings, are shown in the Table 1.

In this experiment, to remove the air and water from the walls of the MGEM chamber, it was baked at 200°C and pumped during 2 hours. After that, the pure commercial neon was used to fill the chamber. During several hours the concentrations of admixtures $H_2O$ and $N_2$ in the neon filling stayed at the level of several ppm, which means that the neon was practically pure. The absence of necessary quantity of quenching admixtures in pure neon results in a large photo effect at the electrodes of the chamber produced by photons, which are emitted by avalanches. This restrict the maximum coefficient of electron multiplication of MGEM to $K_{ampl}^{\max} \cong 10$ for the gas pressure of P=1bar and to $K_{ampl}^{\max} \cong 60$ for P=0.4bar, while the $\beta$-irradiation is used for the ionization of the gas. These results are in a good agreement with previous works [7, 11].

The proportional signals in pure neon at pressures of P=0.4bar and 1.0bar have a rise time, caused by primary avalanches, of Δt≅30μs, and a tail, caused by secondary avalanches, of Δt≅70μs. For the threshold potential difference ΔV=ΔV( $K_{ampl}^{\max}$ ) ± 2V, which corresponds to the maximal amplification of the MGEM, the proportional signals turns into streamers with duration of ~5ms. At further increase of the voltage by several volts, the streamers transform into continuous discharge.

For the mixture *Ne+100* ppm $H_2O$ at pressure of P=1bar and $\beta$-irradiation, the secondary avalanches are delayed relatively to primary ones. When the voltage is increased to the maximum, the amplitudes of the secondary avalanches become several times larger compared to the primary avalanches. The admixture of $H_2O$ in neon decreases the probability of formation of the streamers even at the maximal amplification $K_{ampl}^{\max}$. It is necessary to note that the electrons from MGEM are effectively extracted in the anode gap only at the voltage in anode gap ≥470V, with the corresponding multiplication in the gap $K_{ampl}^{anod}$ ≥66. The reason for such a bad permeability of the lower electrode of the MGEM consists in a low ratio between the diameter of the holes and the thickness of the electrodes (1:1).

The admixture of $N_2$ to neon is a bad quenching addition, because of a large photo effect at the electrodes of the chamber. The reason of that consists probably in presence of meta stable states in $N_2$ molecules with energies of 6.2eV(1.3-2.6s) and 8.4eV(0.5s).

The best results were obtained for the mixtures *Ne*+(1-12)ppm $H_2O$+(10-100)ppm $N_2$, with total pressure of P=1bar. In the range of proportional amplification, the signals with rise time (60-80)μs for these mixtures are observed, and at the threshold voltage of ΔV( $K_{ampl}^{max}$ ) the streamers of broken shape and with duration of (5-10)ms are observed.

# Conclusion

The GEM detectors with wire and metallic electrodes and gas filling in a gap between them have been proposed, produced and investigated. The streamer and spark discharges between the electrodes of the detectors do not destroy these GEM. These GEM do not accumulate the charges between their electrodes and as a result they have high stability. The best gas for these GEM filling is neon with a small admixture of quenching gases, for example *($N_2$+$H_2O$)* at≈100ppm. This filling offers the greatest coefficients of proportional multiplication as compared with other gases, with a small voltage difference between its electrodes and without streamer discharges in the range of proportional amplification.

The GEM with metallic electrodes have high mechanical durability and provide good suppression of the microphone effect, which takes place in the presence of mechanical vibrations. The electrodes of these detectors can be of different, even very complex, shape. For example, such a detector with cylindrical electrodes can be used in tomography equipment.

Based on our investigations and the results of other works, it may be concluded that in Penning mixtures, which are used for GEM filling, the streamers are developed from the secondary avalanches.

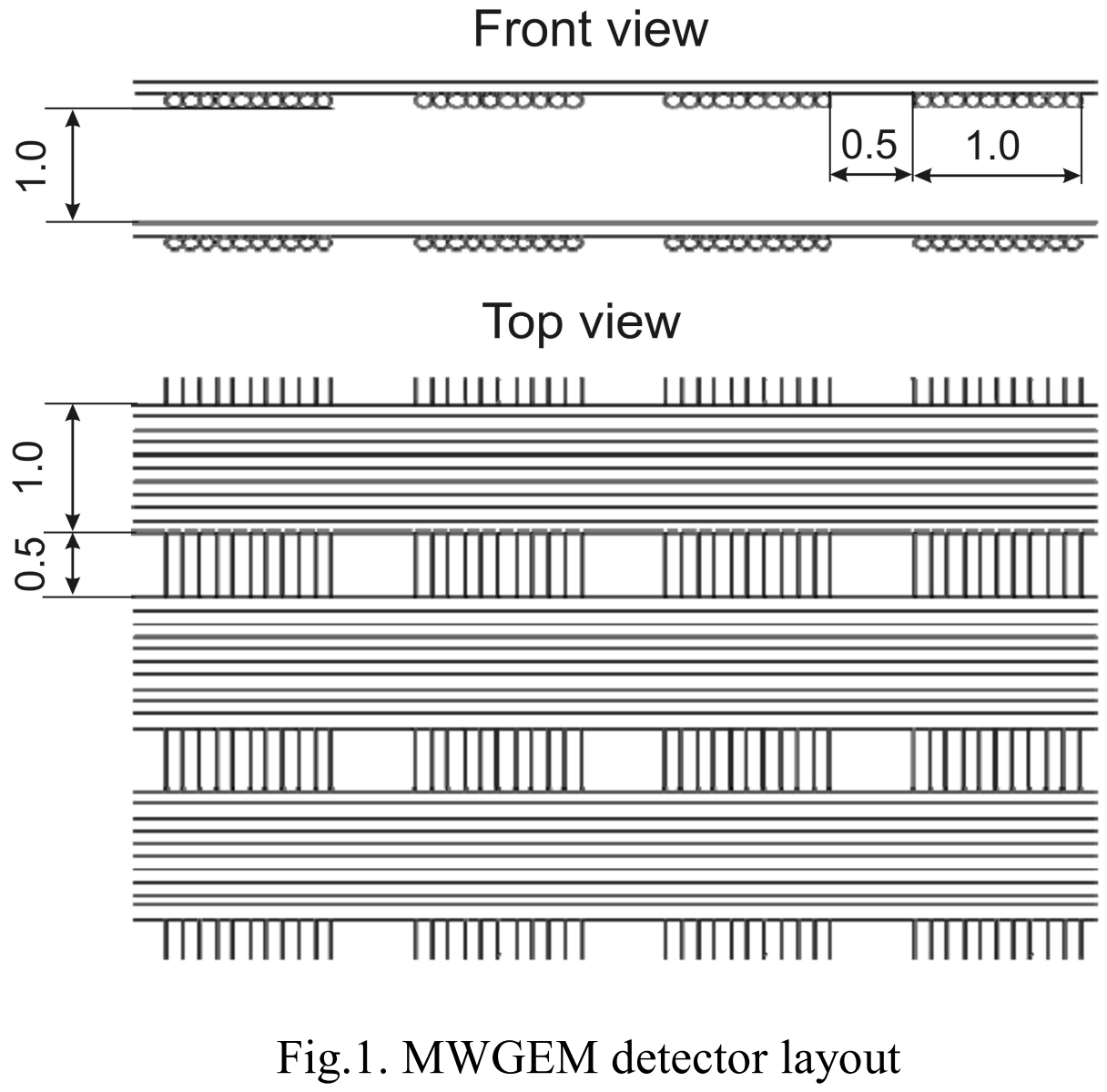

Fig.1. MWGEM detector layout

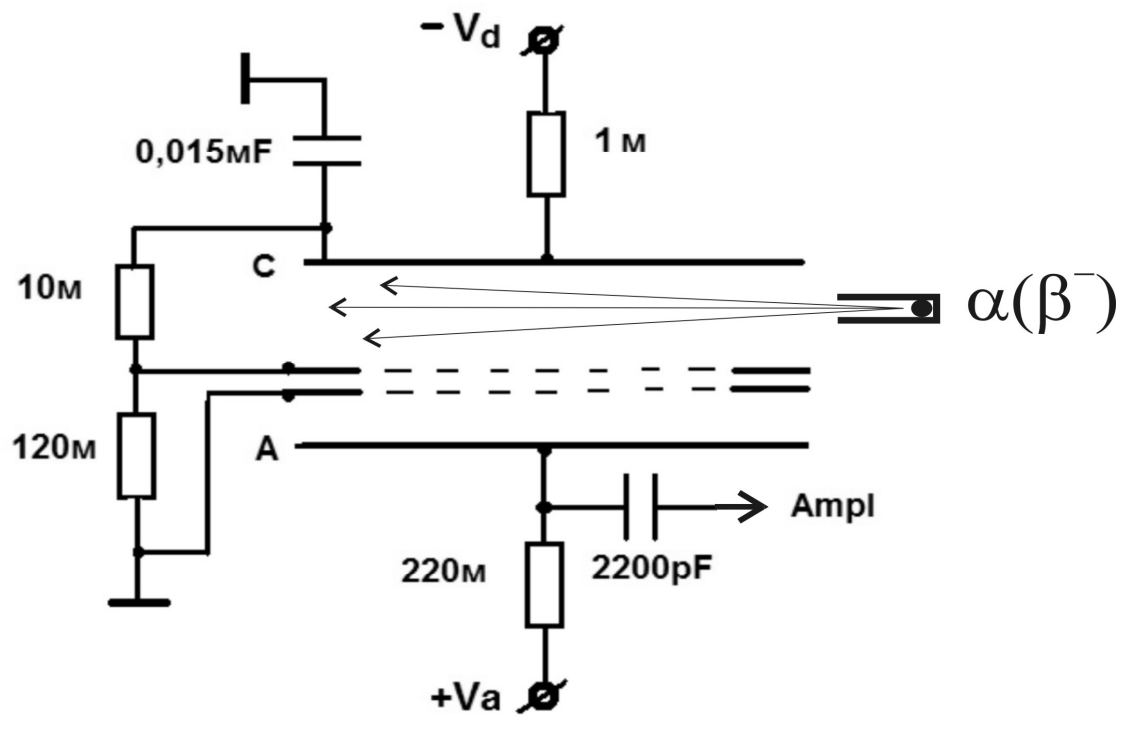

Fig.2. The chamber for MWGEM investigations.

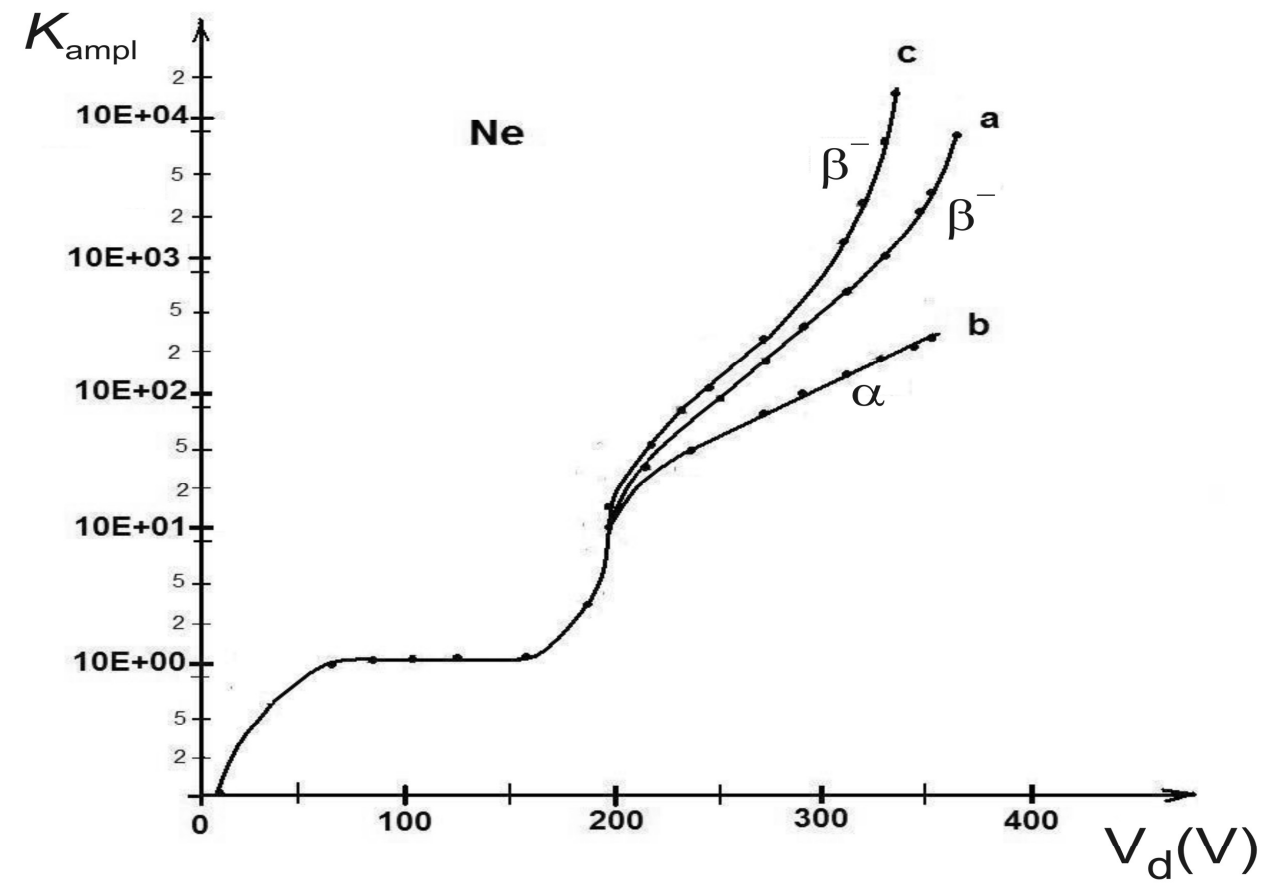

Fig.3. The dependences for the coefficients of proportional electron multiplication from potential difference between GEM electrodes for a gap of 1mm at Ne+(( $H_2O$ +$N_2$)≤100ppm)—filling at 1.0bar.

Curve (a): MWGEM, P=1bar, β-irradiation,

Curve (b): MWGEM, P=1bar, α-irradiation,

Curve (c): MWCAT at P=1bar and β-irradiation.

$V_d$—voltage on resistive dividers of the chambers Fig.2 and Fig.4.

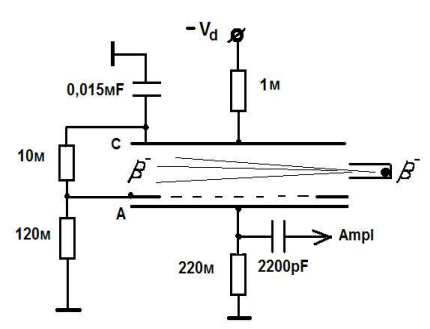

Fig.4. The chamber for MWCAT investigations

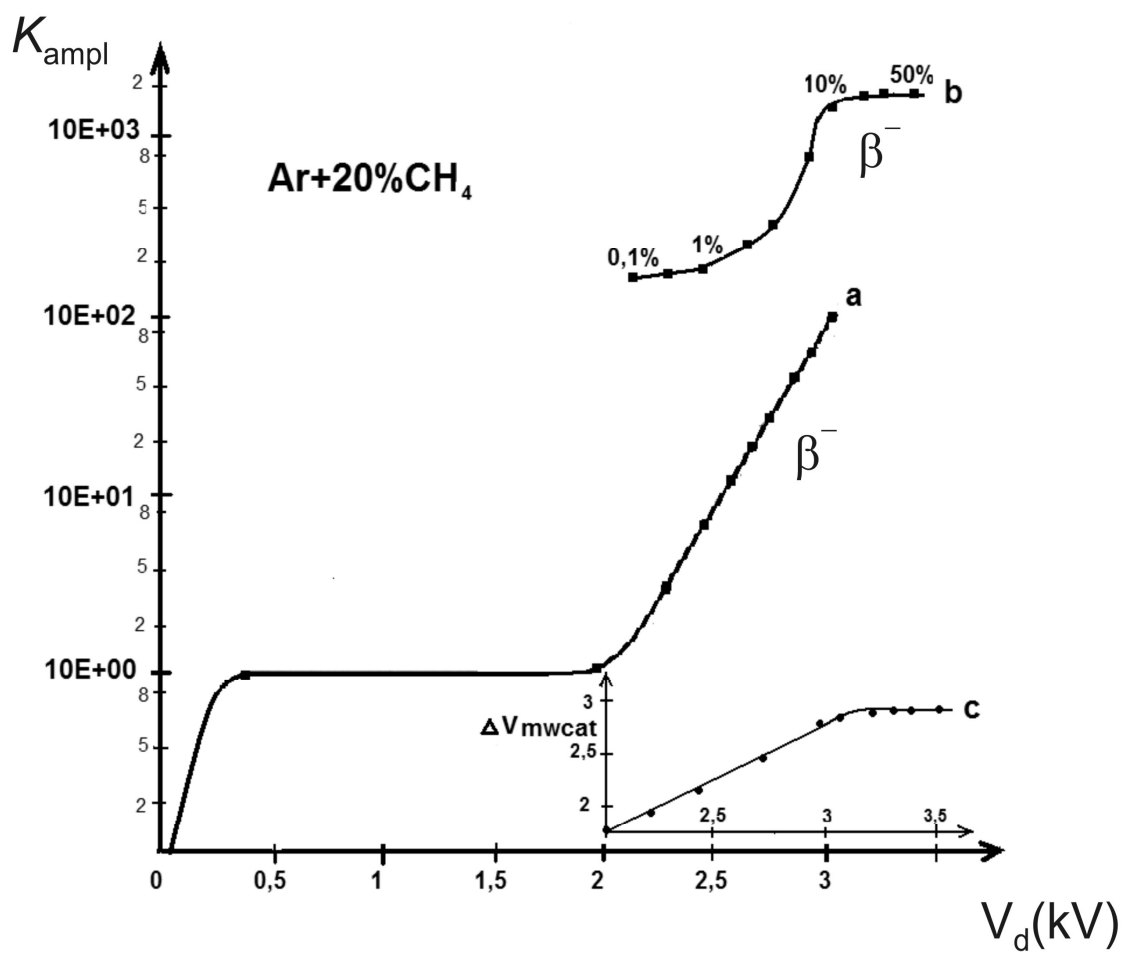

Fig.5. Curve(a):The *Kampl =f (Vd)* of proportional signals for MWCAT at *Ar+20 % $CH_4$* filling and β--irradiation.
Curve (b): *Kampl =f (Vd)* of streamer discharges for MWCAT at *Ar+20 % $CH_4$* filling and $\beta^-$-irradiation. Quantity in% is a part of streamer discharges.
Curve (c): voltage difference between MWCAT electrodes ΔVmwcat as a function of *Vd*.

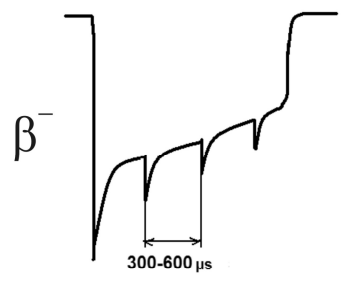

Fig.6. Form of the streamer signal for Ar +20%$CH_4$ mixture and β-irradiation

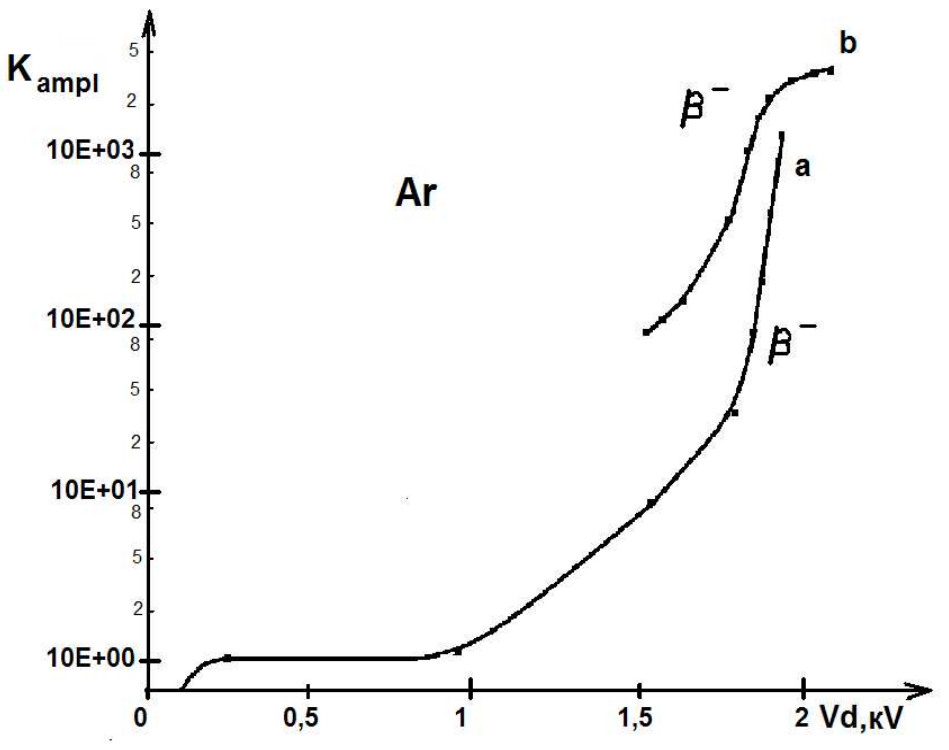

Fig.7. Curve (a): $K_{ampl}$=f($V_d$) of proportional signals for MWCAT at Ar—filling and β--irradiation;

Curve (b): $K_{ampl}$=f($V_d$) of streamer discharges for MWCAT at Ar—filling and β--irradiation.

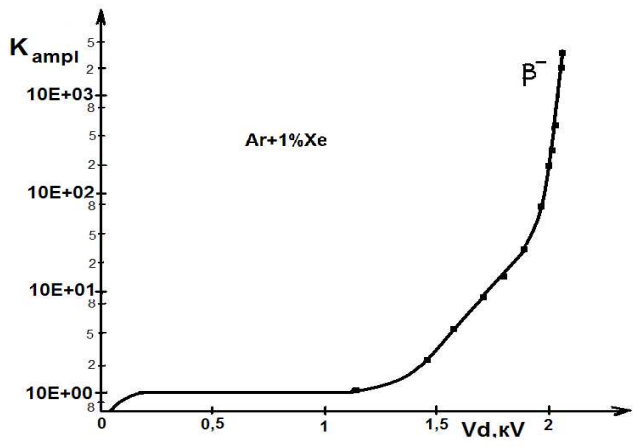

Fig.8. $K_{ampl}=f\,(V_d)$ of proportional signals for MWCAT at *Ar+1% Xe* filling.

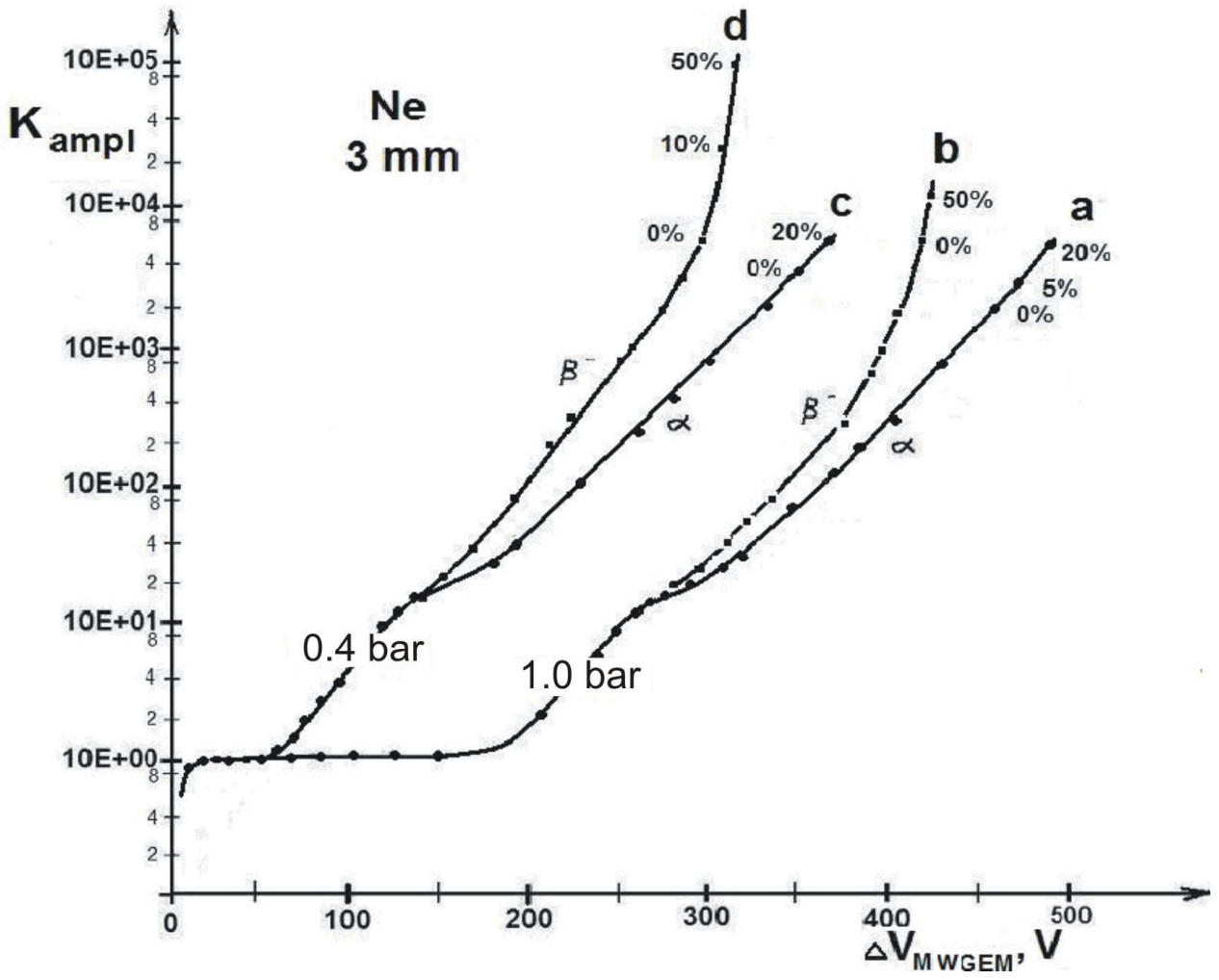

Fig.9. The dependences of the coefficients of proportional electron multiplication on the potentials difference between MWGEM electrodes for a gap 3 mm at pressures 0.4 and 1.0bar:

Curve (a): $\alpha$, P=1.0bar, $K_{ampl}^{\max}=6\cdot10^{3}$ (20% of streamers), no streamers for $K_{ampl}\leq2000$;

Curve (b): $\beta^{-}$, P=1.0bar, $K_{ampl}^{\max}=1.2\cdot10^{4}$ (50% of streamers), no streamers for $K_{ampl}\leq6000$;

Curve (c): $\alpha$, P=0.4bar, $K_{ampl}^{\max}=6\cdot10^{3}$ (20% of streamers), no streamers for $K_{ampl}\leq3.75\cdot10^{3}$;

Curve (d): $\beta^{-}$, P=0.4bar, $K_{ampl}^{\max}=10^{5}$ (50% of streamers), no streamers for $K_{ampl}\leq6\cdot10^{3}$.

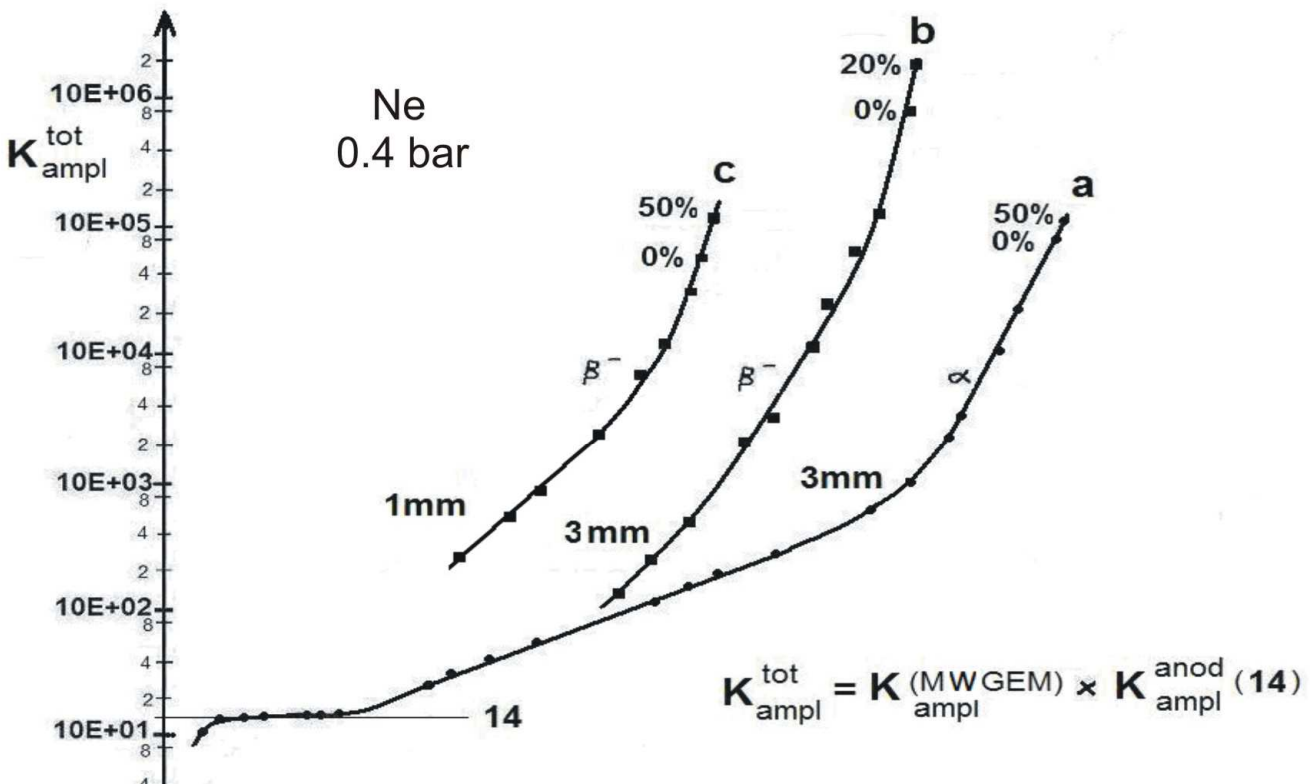

Fig.10. The dependences of the coefficients of proportional electron multiplication

$K_{tot}=K_{ampl}$(MWGEM)×14 on the potentials difference between MWGEM electrodes for gaps 1 and 3mm and neon pressure of 0.4bar:

Curve (a): $\alpha$, 3mm, $K_{tot}^{\max}=1.12\cdot10^{5}$ (50% of streamers), no streamers for $K_{ampl}\leq8\cdot10^{4}$;

Curve (b): $\beta^{-}$, 3mm, $K_{tot}^{\max}=2\cdot10^{6}$ (20% of streamers), no streamers for $K_{ampl}\leq7.8\cdot10^{5}$;

Curve (c): $\beta^{-}$, 1mm, $K_{tot}^{\max}=1.08\cdot10^{5}$ (50% of streamers), no streamers for $K_{ampl}\leq5.9\cdot10^{4}$.

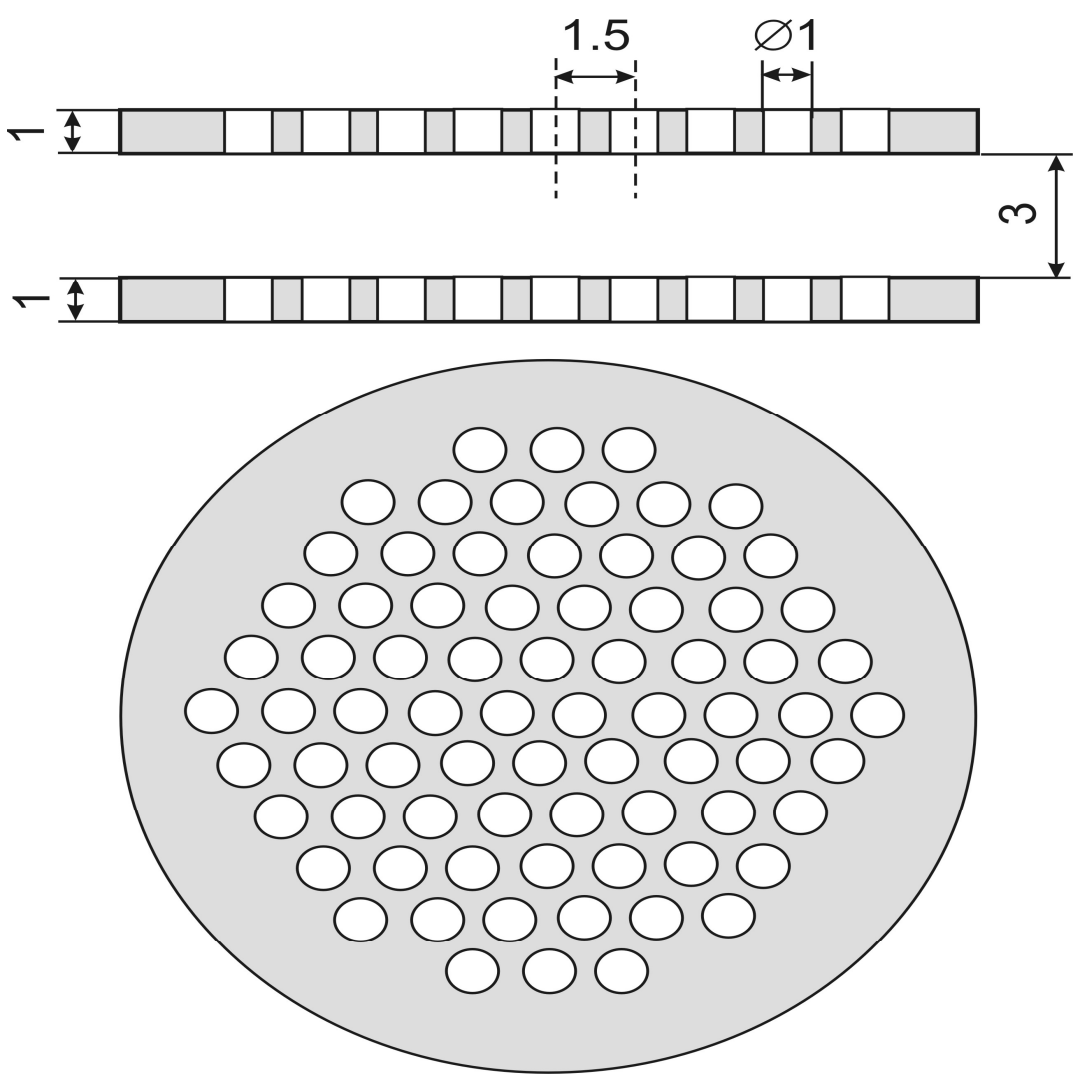

Fig.11. MGEM detector layout

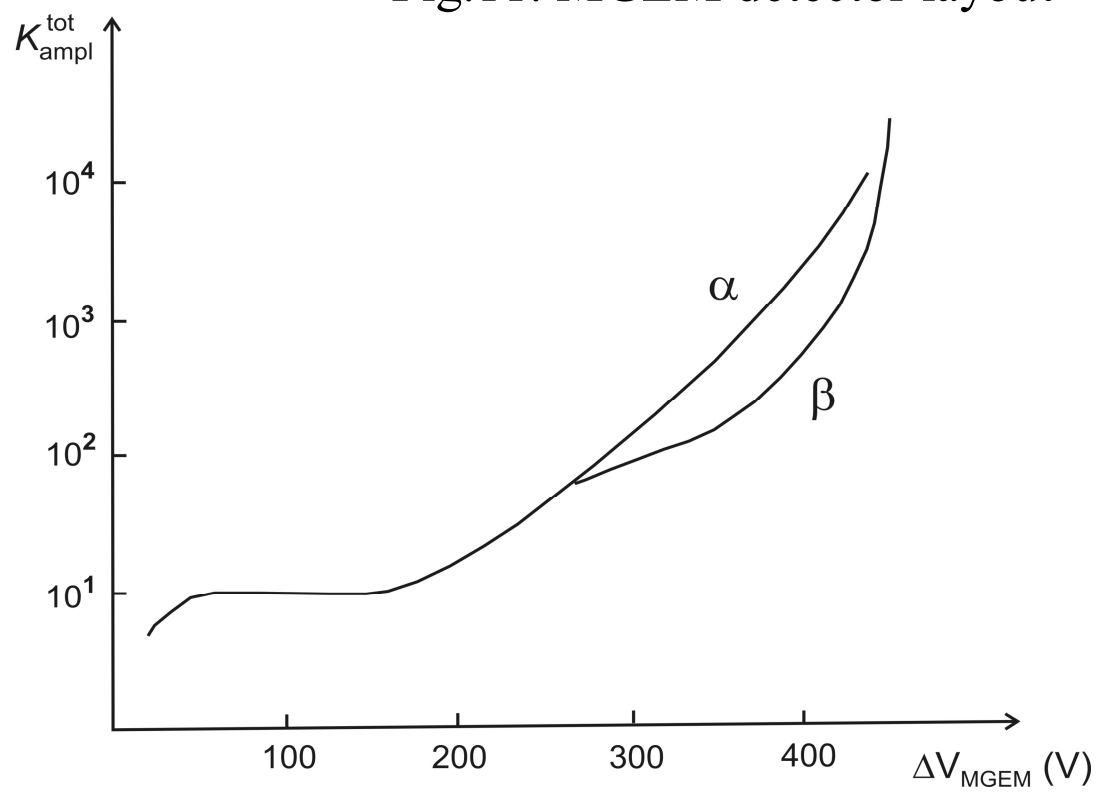

Fig.12. Dependences for the total proportional electron multiplication coefficients $K_{ampl}^{tot}$ from the potential difference between MGEM electrodes at pressure of 1bar and for the amplification of electrons in anode gap, $K_{ampl}^{anod}$=10.

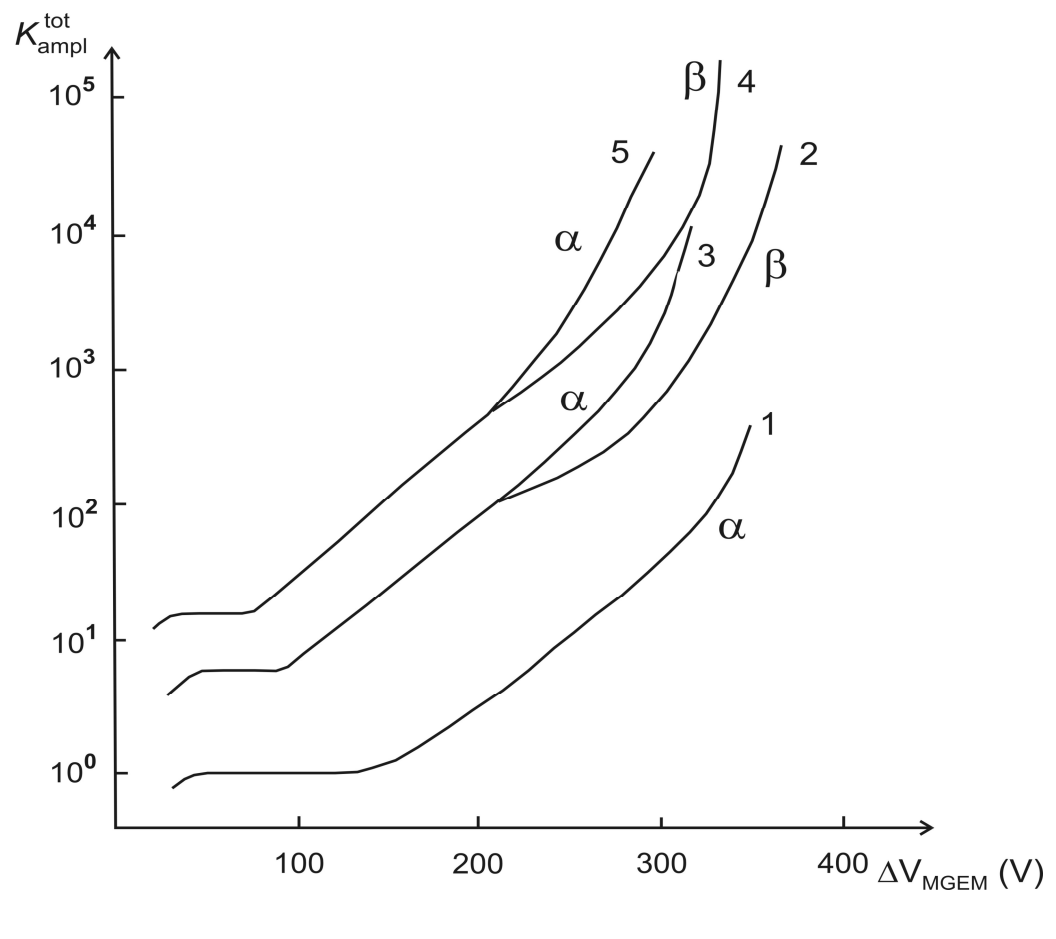

Fig.13. Dependences for the total proportional electron multiplication coefficients $K_{ampl}^{tot}$ from the potential difference between MGEM electrodes at pressure of 0.4bar and for the amplification of electrons in anode gap:

Curve 1 corresponds to $V_a$=250V($K_{ampl}^{anod}$=1);

Curves 2,3 correspond to $V_a$=300V($K_{ampl}^{anod}$=6);

Curves 4,5 correspond to $V_a$=320V($K_{ampl}^{anod}$=15).

<u>Table 1. The signal form for different fillings of chamber at β-irradiation</u>

| Gas content | P(atm) | $+V_A$(V) | $K^{anode}_{ampl}$ | $K^{max}_{ampl}$ | Signal form at proportional region | Signal form at maximal amplification |
|---|---|---|---|---|---|---|
| Pure *Ne* | 1.0 | 400 | 6 | 10 | 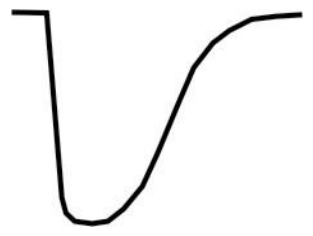 | 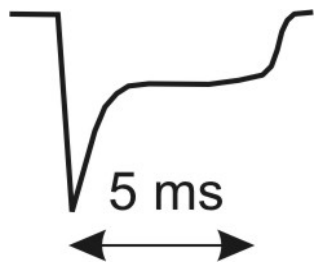  |
| Pure *Ne* | 0.4 | 250 | 1 | 60 | 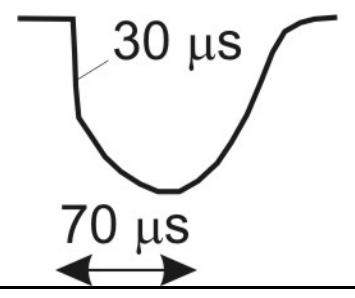  | 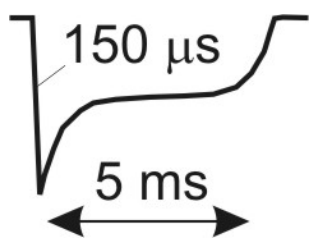  |
| *Ne* + 100ppm $H_2O$ | 1.0 | 470 | 66 | | 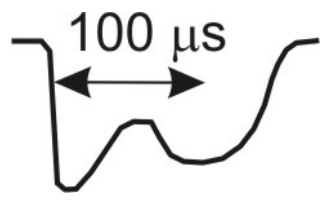  | 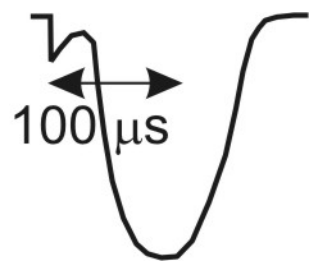  |
| *Ne* + 10ppm $N_2$ | 1.0 | 400 | 6 | | 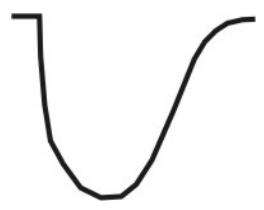 | 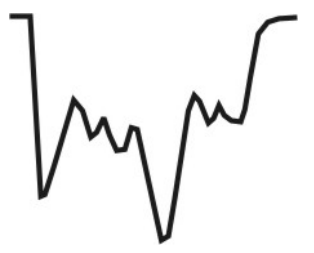 |
| *Ne* + 100ppm $N_2$ | 0.4 | 440 | 10 | | 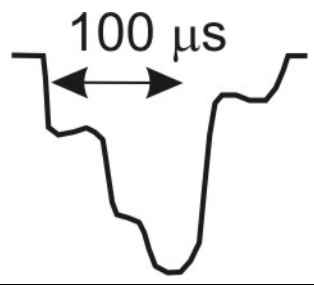  | |
| *Ne* + 200ppm $N_2$ | 1.0 | 400 | 6 | | 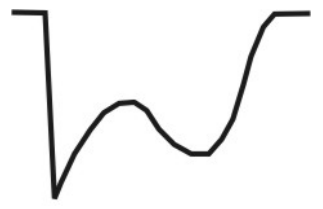 | 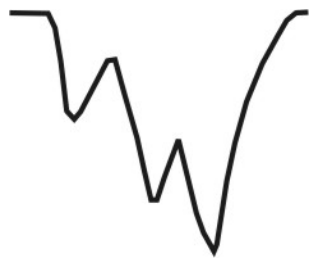 |
| *Ne* + 500ppm $N_2$ | 1.0 | 400 | 6 | | 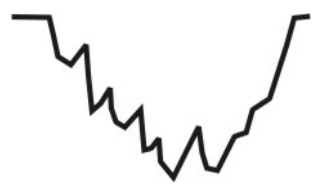 | 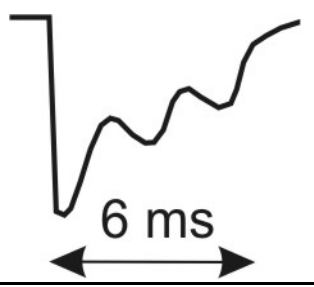  |
| *Ne* + 1ppm $H_2O$ +10ppm $N_2$ | 1.0 | 400 | 6 | | 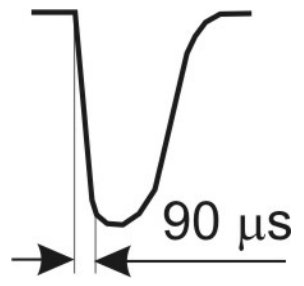  | 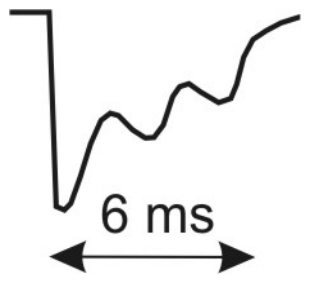  |
| *Ne* + 6ppm $H_2O$ +50ppm $N_2$ | 1.0 | 400 | 6 | | 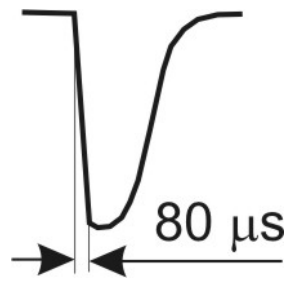  | 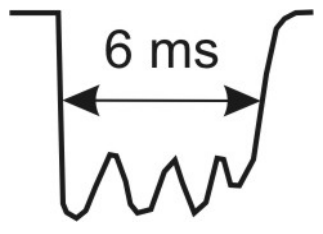  |
| *Ne* + 12ppm $H_2O$ +100ppm $N_2$ | 0.4 | 280 | 4 | | 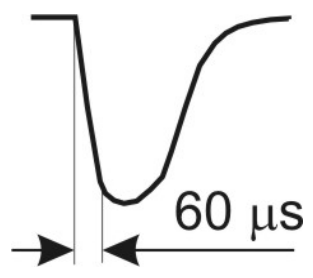  | 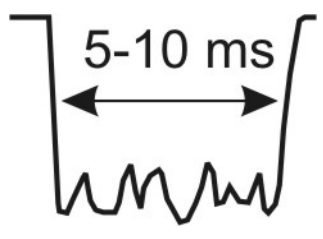  |